\documentclass[sigconf]{acmart}

\AtBeginDocument{%
  \providecommand\BibTeX{{%
    \normalfont B\kern-0.5em{\scshape i\kern-0.25em b}\kern-0.8em\TeX}}}

\setcopyright{acmcopyright}
\copyrightyear{2018}
\acmYear{2018}
\acmDOI{XXXXXXX.XXXXXXX}

\acmConference[KDD '24]{KDD}{Aug 25--29,
  2024}{Barcelona}
\acmPrice{15.00}
\acmISBN{978-1-4503-XXXX-X/18/06}

\usepackage{booktabs}
\usepackage{enumitem}
\usepackage{xcolor}
\usepackage{subfigure}
\usepackage{multicol}
\usepackage{multirow}

\usepackage{algpseudocode}
\usepackage{amsfonts}
 
\usepackage{amssymb}
\usepackage{algorithm}

\usepackage{makecell}
\usepackage{xspace}
\newcommand{\modelname}{DisCo\xspace}

\begin{document}

% \title{Both are Vital: Towards Harmonious Collaboration and Disentanglement between Tabular and Semantic Domain for Sequential CTR Prediction}
\title{DisCo: Towards Harmonious Disentanglement and Collaboration between Tabular and Semantic Space for Recommendation}
%\title{Towards Harmonious Collaboration and Disentanglement between Tabular and Semantic Space for Recommendation}


\author{Kounianhua Du}
\affiliation{%
  \institution{Shanghai Jiao Tong University}
  %\streetaddress{1 Th{\o}rv{\"a}ld Circle}
  \city{Shanghai}
  \country{China}}
\email{kounianhuadu@sjtu.edu.cn}

\author{Jizheng Chen}
\affiliation{%
  \institution{Shanghai Jiao Tong University}
  \city{Shanghai}
  \country{China}}
\email{humihuadechengzhi@sjtu.edu.cn}

\author{Jianghao Lin}
\affiliation{
  \institution{Shanghai Jiao Tong University}
  \city{Shanghai}
  \country{China}}
\email{chiangel@sjtu.edu.cn}

\author{Yunjia Xi}
\affiliation{
  \institution{Shanghai Jiao Tong University}
  \city{Shanghai}
  \country{China}}
\email{xiyunjia@sjtu.edu.cn}

\author{Hangyu Wang}
\affiliation{%
  \institution{Shanghai Jiao Tong University}
  \city{Shanghai}
  \country{China}}
\email{hangyuwang@sjtu.edu.cn}

\author{Xinyi Dai}
\affiliation{%
  \institution{Huawei Noah's Ark Lab}
  \city{Shanghai}
  \country{China}}
\email{daixinyi5@huawei.com}

\author{Bo Chen}
\affiliation{%
  \institution{Huawei Noah's Ark Lab}
  \city{Shanghai}
  \country{China}}
\email{chenbo116@huawei.com}

\author{Ruiming Tang}
\affiliation{%
  \institution{Huawei Noah's Ark Lab}
  \city{Shenzhen}
  \country{China}}
\email{tangruiming@huawei.com}

\author{Weinan Zhang}
\affiliation{%
  \institution{Shanghai Jiao Tong University}
  \city{Shanghai}
  \country{China}}
\email{wnzhang@sjtu.edu.cn}

\renewcommand{\shortauthors}{Kounianhua Du et al.}

\begin{abstract}
Recommender systems play important roles in various applications such as e-commerce, social media, etc. Conventional recommendation methods usually model the collaborative signals within the tabular representation space. Despite the personalization modeling and the efficiency, the latent semantic dependencies are omitted. Methods that introduce semantics into recommendation then emerge, injecting knowledge from the semantic representation space where the general language understanding are compressed. However, existing semantic-enhanced recommendation methods focus on aligning the two spaces, during which the representations of the two spaces tend to get close while the unique patterns are discarded and not well explored.
In this paper, we propose \textbf{\modelname} to \textbf{Dis}entangle the unique patterns from the two representation spaces and \textbf{Co}llaborate the two spaces for recommendation enhancement, where both the specificity and the consistency of the two spaces are captured.
Concretely, we propose 1) a dual-side attentive network to capture the intra-domain patterns and the inter-domain patterns, 2) a sufficiency constraint to preserve the task-relevant information of each representation space and filter out the noise, and 3) a disentanglement constraint to avoid the model from discarding the unique information. These modules strike a balance between disentanglement and collaboration of the two representation spaces to produce informative pattern vectors, which could serve as extra features and be appended to arbitrary recommendation backbones for enhancement.
%For efficient inference, an indexed knowledge base is firstly prepared to pre-store the semantic knowledge vectors after dimension reduction.
Experiment results validate the superiority of our method against different models and the compatibility of \modelname over different backbones. Various ablation studies and efficiency analysis are also conducted to justify each model component. 
\end{abstract}




\received{20 February 2007}
\received[revised]{12 March 2009}
\received[accepted]{5 June 2009}

\maketitle

\section{Introduction}
\label{sec:intro}
%介绍推荐系统重要性及应用
\begin{figure}[t!]
    \centering
    \includegraphics[width=0.48\textwidth, trim=0 0 0 0, clip]{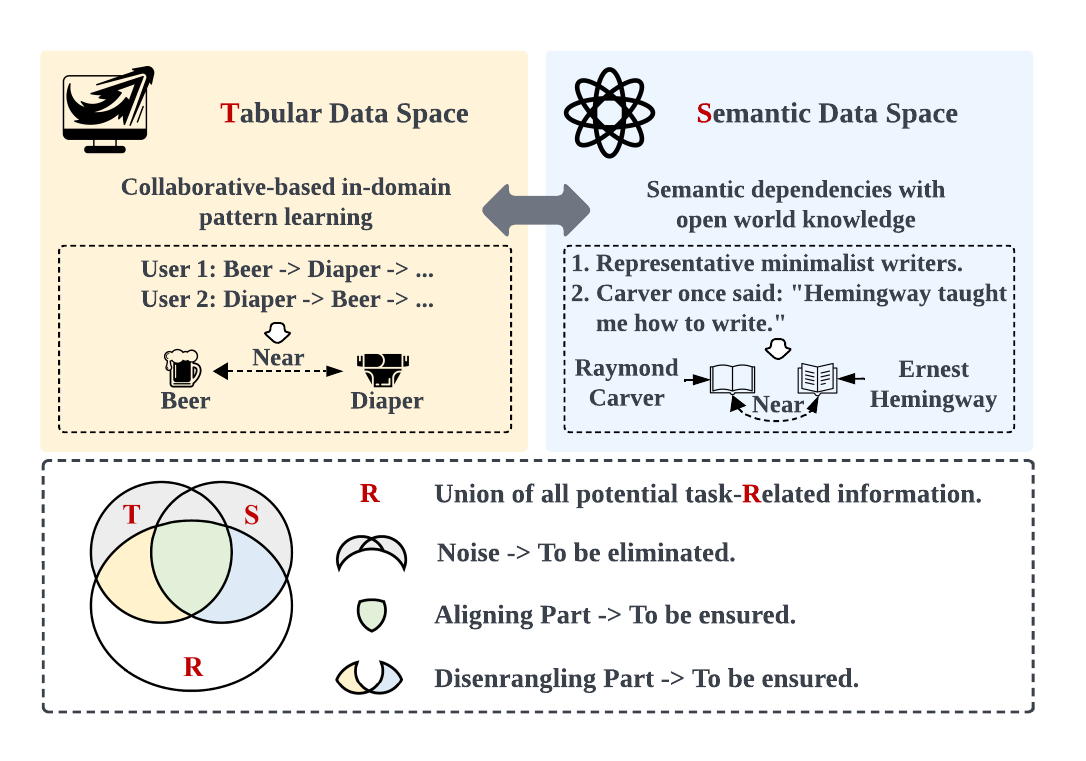}
    \caption{Illustration of the motivation.}
    \vspace{-15pt}
    \label{fig:intro}
\end{figure}
Recommender systems have become an integral part of today's digital ecosystem, enhancing user experiences, boosting engagement, facilitating decision-making, and fostering connections between users and relevant content or products. They are widely used in various industries, including e-commerce \citep{Amazon}, entertainment \citep{entertainment}, social media \citep{news}, and online streaming platforms \citep{18din, 19dien, huang2022neural}.

% 两个space的优劣
Conventional recommender systems usually only model the collaborative signals within the tabular representation space, where samples consist of multi-field categorical features. These methods focus on mining beneficial interactions among features \citep{ffm,afm,autofis,autoint} and modeling user interests using historical user behaviors \citep{18din,19dien,dsin,20sim,20ubr} for accurate and personalized recommendation. While being good at modeling feature interactions and personalized user preferences, these methods fail to learn the latent semantic dependencies of features. For example, as shown in Figure \ref{fig:intro}, a book written by Ernest Hemingway and a book written by Raymond Carver can be close in semantic space because the two authors are known to be minimalist writing style and Carver said that he borrowed many elements from Hemingway's style. This latent semantic dependency cannot be inferred from the tabular representation space where features are firstly encoded in a one-hot manner.
Attempts to incorporate external knowledge within semantic representation space into recommendation then emerge \citep{hou2022universal, li2023ctrl, hou2023learning}, where textual descriptions and their encodings are used to hold the external semantic knowledge. Despite the general language understanding within semantic representation space, the encoding of the large language model has ambiguity and fails to imply the correlations of some features well. For instance, beer and diaper are distant in semantic space but near in tabular space for recommendation where the correlation analysis is done, since users tend to buy beer and diaper together.

% 引入要collaborate两个space
% 现存方法问题：只用一个 or 专注于align，导致disentangled part被discard掉
As discussed above, the relationship among the same set of user behaviors can be different in the two representation spaces.
%, with some of them sharing a similar local structure and the other possessing distinct dependency relations. 
The unique and disentangled patterns from the two representation spaces form a complementary relationship with each other, contributing different information. Therefore, it is vital to effectively and efficiently collaborate the tabular and semantic representation spaces. The existing works either: 1) take only one space into consideration \citep{geng2022recommendation} or 2) only focus on aligning the two spaces \citep{li2023ctrl}, during which the representations tend to get close (green part in Figure~\ref{fig:intro}), with the disentangled part (yellow and blue parts in Figure~\ref{fig:intro}) being discarded.

In this paper, we propose \textbf{\modelname} to \textbf{Dis}entangle and \textbf{Co}llaborate the tabular and semantic representation spaces for user behavior patterns modeling. 
%As shown in Figure~\ref{fig:intro},
%in addition to utilizing the aligning part of the two spaces (green part), we wish to better capture the unique and disentangled information from the two spaces (yellow and blue parts) to complement each other. 
Therefore, we design 1) a dual-side attentive network (DS-Attn) to capture the intra-domain and inter-domain patterns, 2) a sufficiency constraint to preserve the useful information (yellow, green, and blue parts in Figure~\ref{fig:intro}) and eliminate the noisy information (grey part in Figure~\ref{fig:intro}) from the two representation spaces, and 3) a disentanglement constraint to preserve the disentangling parts from the two representation spaces (yellow and blue part in Figure~\ref{fig:intro}). Together, these modules strike a balance of the collaboration and disentanglement between the tabular and semantic representation space. 

Concretely, DS-Attn infers the inner patterns within the tabular space and the semantic space for collaborative-based correlations and semantic-based dependencies with the intra-domain attention, and captures the inter patterns between the two spaces for aligned knowledge with the inter-domain attention where a query-key exchange is adopted to make the two spaces attend to each other. The parameters of DS-Attn and embedding networks are regularized by the sufficiency and disentanglement constraints. The sufficiency constraint maximizes the mutual information between the representations of each space with the labels, which preserves the task-relevant information offered by each space. The disentanglement constraint minimizes the mutual information between the output vectors of DS-Attn from the two spaces, which forces the two spaces to offer different information. The resulting vectors output by DS-Attn and regularized by the two constraints preserve both the consistent and the specific knowledge of the two representation spaces, which can then be fed into arbitrary recommendation backbones for prediction enhancement. 
The main contributions of the paper are summarized as follows:
\begin{itemize}[leftmargin=10pt]
    \item We design a novel and effective framework, \modelname, that harmoniously captures both the consistent and the specific knowledge from tabular representation space and semantic representation space with the dual-side attentive network under the regularization of the proposed sufficiency and disentanglement constraints.
    \item We emphasize the importance of unique and disentangled information in both the tabular space and the semantic space, and propose the first work to disentangle the tabular and semantic representation spaces for unique domain knowledge.
    \item \modelname is a model-agnostic framework compatible with different recommendation backbones, offering flexibility and generality. 
\end{itemize}
The experiment results over a series of recommendation backbones justify the consistent superiority of the proposed method. Ablation studies are also conducted to validate the effectiveness of different model components.
%\Jianghao{in DM, usually experiment serve as the final contribution}
%\xinyi{Stop here}
\section{Related Work}
\label{sec:re}
\subsection{Tabular-Only Methods}
Early recommendation models focus on digging into interactions among features. FM \citep{rendle2010factorization} captures second-order feature interactions. FFM \citep{juan2016field} introduces field-aware interactions. Wide \& Deep \citep{cheng2016wide} combines the strengths of linear models and deep neural networks.
%to leverage both memorization and generalization capabilities.
DeepFM \citep{guo2017deepfm} replaces the logistic regression layer in \cite{cheng2016wide} with an FM layer. xDeepFM \citep{lian2018xdeepfm} introduces the cross layer for high-order feature interactions. PNN \citep{18pnn} utilizes the product layer to learn the high-order product interactions. DCN \citep{dcn} proposes to capture both shallow and deep feature interactions effectively. AutoInt \citep{autoint} utilizes the multi-head self-attention mechanism to learn high-order feature interactions. By modeling user behavior patterns, recommenders provide more personalized recommendations. DIN \citep{18din} incorporates the attention mechanism \citep{vaswani2023attention} to capture user interests. 
%It reduces the influence of historical behavior information that is not relevant to the currently estimated advertisement on the current click estimation judgment. 
DIEN \citep{19dien} %reveals that user interests evolve over time, thus 
uses GRU module to better model the evolving interests of users. 
DSIN \citep{dsin} proposes to capture the dynamic interests of users within a session.
%, enabling personalized recommendations for sequential user interactions. 
 MIMN \citep{mimn} leverages a memory network architecture 
%that can dynamically store and retrieve user interests 
to capture different aspects of user interests. SIM \citep{20sim} models long-term user behaviors.
%and utilizes the SimHash algorithm to aggregate them.

\subsection{Semantic-Enhanced Methods}
Recently, large language models have shown great impact and shed light on various domains of recommendation systems. There are several attempts to incorporate large language models into recommender systems \citep{lin2023recommender,lin2023rella,xi2023openworld,clickprompt}. PTab \citep{liu2022ptab} adopts a classic BERT \citep{devlin2018bert} framework with Modality Transformation(MT), Masked-Language Finetuning(MF), and Classification Fine-tuning(CF) training stages.
%, which enlarges the original training set through mixed tabular datasets. 
P5 \citep{geng2022recommendation}, as well as its variants \citep{geng2023vip5,hua2023up5,hua2023index}, propose to tune T5 \citep{raffel2020exploring} as a unified recommendation model for various downstream tasks
%in a generative manner, and attains great performance improvement on zero-shot tasks. 
ZESRec \citep{ding2021zeroshot} proposes to obtain universal representations from item descriptions through BERT for zero-shot recommendation.
%, and these continuous item indices can naturally generalize to unseen or less-frequent items. 
UniSRec \citep{hou2022universal} learns item representations via a fixed BERT model followed by an MoE-enhanced network.
%and use the representations for cross-domain sequential recommendation. 
CTRL \citep{li2023ctrl} adopts the contrastive learning methodology to align the tabular space and semantic space for recommendation enhancement.
%RecFormer \citep{li2023text} proposes to formulate an item and behaviors as a sequence of sentences, where the model is trained to understand the sentence sequence and retrieve the next sentence. 
%Inspired by TallRec \citep{bao2023tallrec} and TabLLM \citep{hegselmann2023tabllm}, Rella \citep{lin2023rella} pays attention to adapting and empowering a pure large language model for zero-shot and few-shot recommendation tasks, and enhances model performance by retrieval and fine-tuning. 
VQ-Rec \citep{hou2023learning} makes improvements on UnisRec \citep{hou2022universal}, which introduces vector quantization technique.

\section{Preliminaries}
\subsection{Problem Formulation}
The click-through rate (CTR) prediction task aims at accurately predicting the probability of a user clicking an item, which is the core task for recommender systems. Therefore, we mainly focus on the CTR prediction task in this work.
The conventional CTR prediction task within the tabular domain can be formulated as 
\begin{equation}
    p(y_i|\mathbf{X^U_i, X^I_i, X^C_i},\theta),
\end{equation}
where $\mathbf{X^U_i}=\left[x_{i,1}^U, \dots, x_{i,F_U}^U\right]$
is the set of user features, $\mathbf{X^I_i}=\left[x_{i,1}^I, \dots, x_{i,F_I}^I\right]$
is the set of item features, $\mathbf{X^C_i}=\left[x_{i,1}^C, \dots, x_{i,F_C}^C\right]$
is the set of context features for the click prediction event (e.g., device, season, etc.), $y_i$ is the label of the data point, and $\theta$ is the model parameter. We use $F_U, F_I, F_C$ to denote the number of user features, item features, and context features, respectively. These methods only focus on modeling feature interactions of the input based on the target sample only but fail to model the user behavior patterns.

Modeling user behavior patterns plays an important role in boosting personalized recommendation performance. This line of methods takes users' historical behaviors as extra inputs and models the dependencies between the candidate item and historical items, which can be formulated as 
\begin{equation}
    p(y_i|\mathbf{X^U_i, X^I_i, X^C_i}, [\langle \mathbf{X^I_{i_k}}, y_{i_k}\rangle]_{k=1}^K, \theta),
\end{equation}
where $[\langle \mathbf{X^I_{i_k}}, y_{i_k}\rangle]_{k=1}^K$ represents the list of user's historical behaviors and their corresponding labels, and $\theta$ is the model parameter. 

%To better model the relations between user behaviors, we could further involve in the negative histories and the label information for prediction enhancement, which can be formulated as
%\begin{equation}
%    p(y|X^U_u, X^I_i, X^C_c, \{\langle X^I_{i_k}, y_{i_k}\rangle\}_{k=1}^K, \theta),
%\end{equation}
%where $[\langle X^I_{i_k}, y_{i_k}\rangle]_{k=1}^K$ represents the list of user's historical behaviors and their corresponding rating information or click-or-not signals.

In this paper, we aim to involve the embedded open-world knowledge of large language models to achieve harmonious disentanglement and collaboration between the tabular and semantic space for recommendation enhancement. Hence, the prediction can be formulated as 
%\begin{equation}
%    p(y_i|\mathbf{X^U_i, X^I_i}, \Phi_S(T_{i}^I), \mathbf{X^C_i}, [\langle \mathbf{X^I_{i_k}}, \Phi_S(T_{i_k}^I),y_{i_k}\rangle]_{k=1}^K, \theta),
%\end{equation}
\begin{equation}
    p(y_i|\mathbf{X^U_i, X^I_i},\mathbf{X^C_i}, [\langle \mathbf{X^I_{i_k}},y_{i_k}\rangle]_{k=1}^K, \Phi_S, \theta),
\end{equation}
where $\Phi_S$ represents the encoder of a large language model.
%\Jianghao{suggest use `large`, not `pretrained`}.

\subsection{Mutual Information Minimization \& Maximization}
Mutual information (MI) is important but hard to compute in neural networks. For mutual information maximization, MINE \citep{mine} builds connections between the expectations of variables and mutual information, and proposes a lower bound of the mutual information based on the Donsker-Varadhan representation of KL divergence.

DIM \citep{dim} points out that we do not necessarily need to obtain the precise value of MI but only need to maximize it. They use the Jensen-Shannon Divergence to estimate the MI and therefore propose a GAN-style loss to maximize it:
\begin{equation}
    L=E_{\mathbb{J}}\left[\log T_w(x,y)\right] + E_{\mathbb{M}}\left[\log(1-T_w(x,y))\right].
\end{equation}

For mutual information minimization, CLUB \citep{club} introduces an upper bound for mutual information. When the conditional distribution  $p(y|x)$ is known, the upper bound can be represented as
\begin{equation}
\begin{split}
    I(X;Y)&\leq I_{CLUB}(X;Y)\\
    &= E_{p(x,y)}\left[\log p(y|x)\right]-E_{p(x)}E_{p(y)}\left[\log p(y|x)\right].
\end{split}
\end{equation}
When the conditional distribution is not known, one could use a variational distribution $q_\theta(y|x)$ to approximate it and the upper bound then becomes
\begin{equation}
    I_{vCLUB}=E_{p(x,y)}\left[\log q_\theta(y|x)\right]-E_{p(x)}E_{p(y)}\left[\log q_\theta(y|x)\right].
\end{equation}

\section{Methodology}
In this section, we describe the methodology of \modelname, which is model-agnostic and compatible with different backbones. 

\subsection{Overview}
\begin{figure*}[!t]
    \centering
    \includegraphics[width=1.0\textwidth]{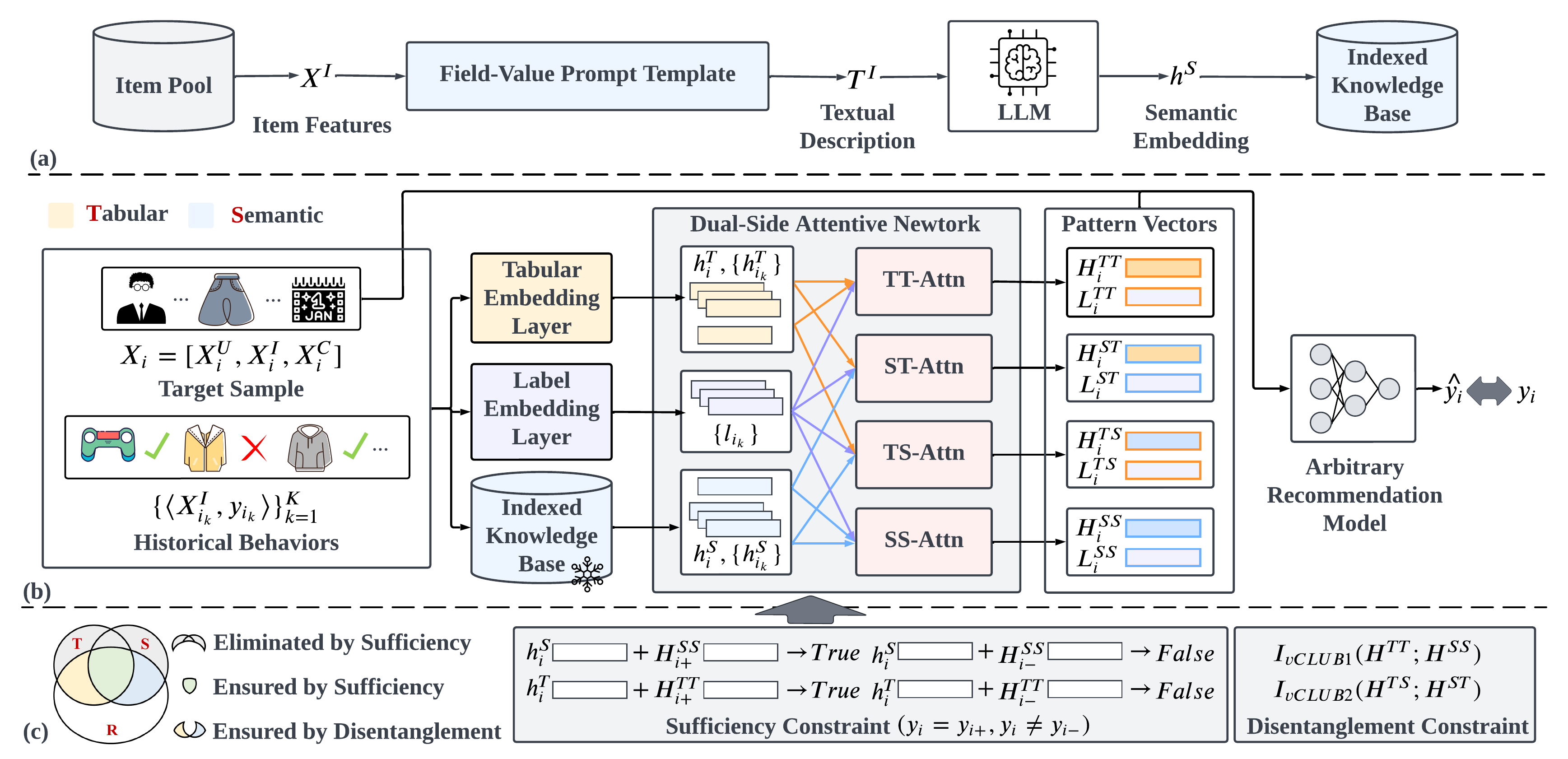}
    \caption{Overview. (a) To extract the semantic knowledge, a textual description is obtained for each item using a field-value prompt template, which is then fed to a LLM for semantic embedding and stored in an indexed knowledge base. (b) The candidate item and the historical behaviors are encoded in tabular and semantic representation spaces, which are then sent to Dual-Side Attentive Network for intra-domain and inter-domain pattern vectors. The resulting pattern vectors serve as extra features and can be appended to arbitrary recommendation model. (c) Two constraints are devised to regularize the model and preserve both the aligning part and the disentangling part of useful information from the two representation spaces. The sufficiency constraint is applied on the behavior vectors and the summarized pattern vectors to preserve the useful information. The disentanglement constraint is applied on the pattern vectors from the two different domains to force the model to capture unique information from both domains.}
    \label{fig:overview}
    %\vspace{-12pt}
\end{figure*}

As illustrated in Figure~\ref{fig:overview}, we first prepare the tabular and semantic embeddings. In this paper, we encode the user behaviors in tabular and semantic representation space and extract patterns from the resulting tabular and semantic embeddings. To make use of the general open-world knowledge from the semantic space in an efficient manner, we pre-store the semantic embedding of each item in an indexed knowledge base, with the item ID being the indexing key. The semantic embeddings are generated from a frozen LLM, with the textual item descriptions as inputs. The tabular embeddings are obtained via tabular embedding layers with one-hot encoded features as inputs.

The proposed \modelname mainly consists of three components: the dual-side attentive network, the sufficiency constraint, and the disentanglement constraint. To collaborate the tabular and semantic representation space, we propose a \emph{Dual-Side Attentive Network} to encode the patterns from user behaviors under collaborative learning based representation space, semantic dependencies-based representation space, and the interactions of the two spaces. The resulting pattern vectors serve as additional features for arbitrary recommendation models.
In addition, we propose two constraints to regularize the model, which preserve the useful and unique information from the two representation spaces: 1) A sufficiency constraint to maximize the mutual information between encoded vectors and the labels, which preserves the task related information and eliminates the noise. 2) A disentanglement constraint to minimize the mutual information between vectors from different representation spaces, which forces each space to provide unique domain-specific knowledge. 
Together, these components strike a harmonious balance between collaboration and disentanglement.

\subsection{Indexed Knowledge Base}
To efficiently utilize the semantic knowledge, we first build an indexed knowledge base to extract and pre-store the semantic embedding of each item. 

%\vspace{-10pt}
\begin{figure}[h]
    \centering
    \includegraphics[width=0.45\textwidth]{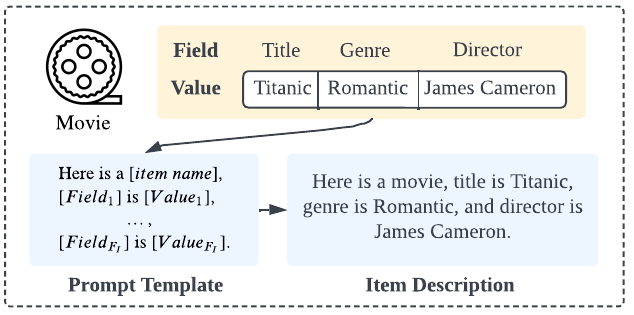}
    \vspace{-10pt}
    \caption{The illustration of the field-value prompt template.}
    \label{fig:kg}
\end{figure}
%\vspace{-10pt}
For each item, we obtain a semantic description for it using the field-value prompt template. As shown in Figure~\ref{fig:kg}, for a movie with title \textit{Titanic}, genre \textit{Romantic}, and director \textit{James} \textit{Cameron}, we can obtain the item description \textit{"Here is a movie, title is Titanic, genre is Romantic, and director is James Cameron."}.
%\begin{equation}
%    \text{\quad} X_i^I \xrightarrow[]{\text{Field-Value Prompt Template}} T_i.
%    \label{eq:prompt}
%\end{equation}
The item description is then fed into a large language model $\Phi_S(\cdot)$ to acquire the semantic embedding, which will be stored in an indexed knowledge base $KB[\cdot]$ for further usage, with the item features being the index key and the semantic embedding being the value. 

%The item description is then fed into a large language model with encoder $\Phi_S(\cdot)$\ljh{into a large language model $\Phi_S(\cdot)$} and stored in an indexed knowledge base $KB[\cdot]$ for further usage, with the item features\ljh{item ID?} being the index key and the item description embedding being the value. 
%This process can be formulated as
%\begin{align}
%    \mathbf{h_i^S} = &\Phi_S(T_i),\label{eq:kgenc}\\
%    KB&[i] = \mathbf{h_i^S},
%    \label{eq:kgstore}
%\end{align}
%where $\Phi_S(\cdot)$ denotes the pretrained large language model, $\mathbf{h_i^S}$ denotes the resulting item semantic knowledge vector, and $KB$ denotes the indexed knowledge base. 
%\vspace{-5pt}
\subsection{Dual-Side Attentive Network}
As discussed in Section~\ref{sec:intro}, the relations among the same set of user behaviors can be different in the tabular and semantic domains. In order to capture the distinct and shared patterns among the behaviors in both domains, we design a dual-side attentive network (DS-Attn) module, which consists of intra-domain attention and inter-domain attention. The intra-domain attention models the distinct domain-specific patterns within the tabular domain for collaborative-based correlations and within the semantic domain for semantic-based dependencies, respectively. The inter-domain attention models the shared patterns between the two domains, where a query-key exchange between the two domains is adopted to make the two domains attend to each other.
%\vspace{-5pt}
\subsubsection{Behaviors Encoding}
For each target sample $\mathbf{X_i=[X_i^U, X_i^I, X_i^C]}$, we gather $K$ recent historical behaviors $\{\langle \mathbf{X_{i_k}^I}, y_{i_k}\rangle\}_{k=1}^K$ to assist the prediction. The candidate item and the historical items are transformed into representations in tabular domain and semantic domain as follows, $\forall j\in\{i,i_1,\dots,i_K\}$:
\begin{align}
\mathbf{h_j^S} &= MLP(KB[\mathbf{X_j^I}]), \label{eq:senc}\\
\mathbf{h_j^T} &=\Phi_T(\mathbf{X_j^I}),\label{eq:tenc}
\end{align}
where $\Phi_T$ denotes the tabular domain embedding network and $KB[\cdot]$  denotes the indexed knowledge base constructed in the previous stage. $MLP$ is used to reduce the dimension of the semantic embedding. $\mathbf{h_j^S}$ and $\mathbf{h_j^T}$ denote the embedding of item features in semantic space and tabular space, respectively. In order to decrease the entanglement of the embeddings for intra-domain and inter-domain pattern extraction, we decouple the corresponding embeddings into chunks. $\forall j\in\{i,i_1,\dots,i_K\}$:
\begin{align}
    \mathbf{h_j^S} &= [\mathbf{h_j^{SI}, h_j^{SC}}],\\
    \mathbf{h_j^T} &= [\mathbf{h_j^{TI}, h_j^{TC}}],
\end{align}
where $\mathbf{h_j^{SI}}\in R^{\frac{d}{2}}, \mathbf{h_j^{SC}}\in R^{\frac{d}{2}}, \mathbf{h_j^{TI}}\in R^{\frac{d}{2}},$ and $\mathbf{h_j^{TC}}\in R^{\frac{d}{2}}$.

In addition, following~\citep{21rim, du2022learning}, we also encode the historical labels to model accurate click signals, $\forall k\in\{1,\dots,K\}$:
\begin{equation}
    \mathbf{l_{i_k}} = \Phi_Y(y_{i_k}),
\end{equation}
where $\Phi_Y$  denotes the label embedding network.

\subsubsection{Attention Block of DS-Attn}
The dual-side attentive network consists of four attention blocks for tabular, semantic, tabular-to-semantic, and semantic-to-tabular user behavior patterns. We first define the function of the basic attention block as follows:
\begin{align}
    \mathbf{H, L'} = Attn_\theta(\mathbf{Q,K,V,L}),\\
    \theta = \{\mathbf{W_Q, W_K, W_V}\},
\end{align}
where $\theta$ denotes the parameters of an attention block. The detail operations for $\mathbf{H, L'} = Attn_\theta(\mathbf{Q,K,V,L})$ are: 
\begin{align}
    \mathbf{Q'} = \mathbf{QW_Q}, \text{ }
    \mathbf{K'} &= \mathbf{KW_K},\text{ }
    \mathbf{V'} = \mathbf{VW_V},\\
    \mathbf{A} = soft&max(\frac{\mathbf{Q'K'^T}}{\sqrt{d}}),\\
    \mathbf{H} = \mathbf{A}\mathbf{V'}&, \text{ }\mathbf{L'} = \mathbf{AL},
\end{align}
 where $\mathbf{Q}\in R^{1\times \frac{d}{2}}, \mathbf{K}\in R^{K\times\frac{d}{2}},\mathbf{V}\in R^{K\times\frac{d}{2}}, \mathbf{L}\in R^{K\times d}$ denote the input for query, key, value, and labels, and $\mathbf{W_Q, W_K, W_V} \in R^{\frac{d}{2}\times d}$ represent the trainable weights.
 
\subsubsection{Intra-Domain Attention}

%\dk{Todo: Too many redundant equations. Define an attention operator and directly use it here.}
To capture the inner domain tabular and semantic behavior patterns, we apply the intra-domain attention among the candidate item and the behavior sequences within the tabular and semantic domains.

We define the query, key, and value for intra-tabular and intra-semantic behavior patterns as
\begin{align}
    \mathbf{Q^{SS}_i} = \mathbf{h^{SI}_i}, &\text{\quad} \mathbf{Q^{TT}_i} = \mathbf{h^{TI}_i}\\
    \mathbf{K^{SS}_i=V^{SS}_i}= \left(       
  \begin{array}{c}   
    \mathbf{h^{SI}_{i_1}} \\  
    \cdots\\  
    \mathbf{h^{SI}_{i_K}}
  \end{array}
\right), &\text{\quad} \mathbf{K^{TT}_i=V^{TT}_i}= \left(       
  \begin{array}{c}   
    \mathbf{h^{TI}_{i_1}} \\  
    \cdots\\  
    \mathbf{h^{TI}_{i_K}}
  \end{array}
\right).
\end{align}
where $\mathbf{Q^{SS}_i}$, $\mathbf{K^{SS}_i}$, and $\mathbf{V^{SS}_i}$ denote the input for intra-semantic attention and $\mathbf{Q^{TT}_i}$, $\mathbf{K^{TT}_i}$, and $\mathbf{V^{TT}_i}$ represent the input of intra-tabular attention.

Then the user behavior patterns in different spaces can be obtained by
\begin{align}
    \mathbf{P^{SS}_i= [H^{SS}_i, L^{SS}_i]} &= Attn_{\theta_{SS}}(\mathbf{Q^{SS}_i, K^{SS}_i, V^{SS}_i,L_i}),\\
    \mathbf{P^{TT}_i = [H^{TT}_i, L^{TT}_i]} &= Attn_{\theta_{TT}}(\mathbf{Q^{TT}_i, K^{TT}_i, V^{TT}_i, L_i}).
\end{align}
%Each intra-domain representation is transformed into query, key, and value as follows:
%\begin{align}
%\mathbf{Q^{TI}_i} &= \mathbf{h^{TI}_iW^{TI}_Q}, \\
%\forall j\in\{1,\dots, K\}: \text{ }\mathbf{K^{TI}_{ij}} &= \mathbf{h^{TI}_{ij}W^{TI}_K}, \text{ }\mathbf{V^{TI}_{ij}} = \mathbf{h^{TI}_{ij}W^{TI}_V}, \\
%\mathbf{Q^{SI}_i} &= \mathbf{h^{SI}_iW^{SI}_Q}, \\
%\forall j\in\{1,\dots, K\}: \text{ }\mathbf{K_{ij}^{SI}} &= \mathbf{h^{SI}_{ij}W^{SI}_K}, \text{ }\mathbf{V_{ij}^{SI}} = \mathbf{h^{SI}_{ij}W^{SI}_V},
%\end{align}
%where $\mathbf{W^{TI}_{Q},W^{TI}_{K},W^{TI}_{V}}$ represent the weights used to transform tabular representations, and $\mathbf{W^{SI}_{Q},W^{SI}_{K},W^{SI}_{V}}$ represent the weights used to transform semantic representations.

%Then we can obtain the distance relationships in each space measured by attention values as follows:
%\begin{align}
%    \mathbf{\alpha^{SS}_{ij}} &= softmax_{j\in\{1, \dots, K\}}\left(\frac{\mathbf{Q^{SI}_iK^{SI}_{ij}}}{\sqrt{d}}\right), \\
%    \mathbf{\alpha^{TT}_{ij}} &= softmax_{j\in\{1, \dots, K\}}\left(\frac{\mathbf{Q^{TI}_iK^{TI}_{ij}}}{\sqrt{d}}\right).
%\end{align}

%The user behavior dependencies in different spaces can then be obtained by
%\begin{align}
%\mathbf{H^{TT}_i} = [\sum_{j\in\{1,\dots,K\}}\mathbf{\alpha^{TI}_{ij}V^{TI}_{ij}}, \sum_{j\in\{1,\dots,K\}}\mathbf{\alpha^{TI}_{ij}l_{ij}}]\\
%\mathbf{H^{SS}_i} = [\sum_{j\in\{1,\dots,K\}}\mathbf{\alpha^{SI}_{ij}V^{SI}_{ij}}, \sum_{j\in\{1,\dots,K\}}\mathbf{\alpha^{SI}_{ij}l_{ij}}]
%\end{align}

\subsubsection{Inter-Domain Attention}
We further capture the inter-domain patterns and model the shared knowledge with an inter-domain attention module.

We define the query, key, and value for intra-tabular and intra-semantic behavior patterns encoded as
\begin{align}
    \mathbf{Q^{ST}_i = h^{SC}_i}, &\text{\quad} \mathbf{Q^{TS}_i = h^{TC}_i}\\
    \mathbf{K^{ST}_i=V^{ST}_i}= \left(       
  \begin{array}{c}   
    \mathbf{h^{TC}_{i_1}} \\  
    \cdots\\  
    \mathbf{h^{TC}_{i_K}}
  \end{array}
\right), &\text{\quad} \mathbf{K^{TS}_i=V^{TS}_i}= \left(       
  \begin{array}{c}   
    \mathbf{h^{SC}_{i_1}} \\  
    \cdots\\  
    \mathbf{h^{SC}_{i_K}}
  \end{array}
\right).
\end{align}

Then the user behavior patterns in different spaces can be obtained by
\begin{align}
    \mathbf{P^{ST}_i=[H^{ST}_i, L^{ST}_i]} &= Attn_{\theta_{ST}}(\mathbf{Q^{ST}_i, K^{ST}_i, V^{ST}_i,L_i}),\\
    \mathbf{P^{TS}_i=[H^{TS}_i, L^{TS}_i]} &= Attn_{\theta_{TS}}(\mathbf{Q^{TS}_i, K^{TS}_i, V^{TS}_i, L_i}).
\end{align}

\subsection{Sufficiency Constraint}
To preserve the task-related information from the two representation spaces and filter out the noises, we maximize the mutual information between the encoded vectors $\mathbf{h}$ from each domain and the labels. The difficulty here lies in that the label space is discrete and only consists of two values. Therefore, we follow DIM \citep{dim} to use the summarized pattern vectors $\mathbf{H_{\cdot}}$ from each space to represent the label space, which extends the discrete states into high-dimensional continuous space. To maximize the mutual information, we maximize the distance between the marginal distribution and joint distributions between two variables.

Concretely, for tabular domain sufficiency preserving, we sample the positive pairs as $\mathcal{I^{T+}}=\{\langle\mathbf{h_i^{TI}, H_{i^+}^{TT}}\rangle| y_i=y_{i+}\}$ and sample the negative pairs as $\mathcal{I^{T-}}=\{\langle\mathbf{h_i^{TI}, H_{i^-}^{TT}}\rangle| y_i\neq y_{i-}\}$. For semantic domain sufficiency preserving, we sample the positive pairs as $\mathcal{I^{S+}}=\{\langle\mathbf{h_i^{SI}, H_{i^+}^{SS}}\rangle| y_i=y_{i+}\}$ and sample the negative pairs as $\mathcal{I^{S-}}=\{\langle\mathbf{h_i^{SI}, H_{i^-}^{SS}}\rangle| y_i\neq y_{i-}\}$.
%Concretely, we sample behavior vectors $\mathbf{h}$ and the summarized pattern vectors $\mathbf{H}$ of the same class as the positive pairs, and we sample behavior vectors $\mathbf{h}$ and the summarized pattern vectors $\mathbf{H}$ of the opposite class as the negative pairs. 
The discriminator network $\mathcal{D}_{\theta 1}, \mathcal{D}_{\theta 2}$ are used to distinguish the two pairs. The optimization objective is:
\begin{equation}
\begin{split}
    l_{Suf} = -\sum_{\mathcal{I^{T+}},\mathcal{I^{T-}}} \left( \log \mathcal{D}_{\theta 1}(\mathbf{h^{TI}_i}, \mathbf{H^{TT}_{i+}})  + \left( 1-\log \mathcal{D}_{\theta 1}(\mathbf{h^{TI}_i}, \mathbf{H^{TT}_{i-}})\right) \right) \\-\sum_{\mathcal{I^{S+}},\mathcal{I^{S-}}} \left( \log \mathcal{D}_{\theta 2}(\mathbf{h^{SI}_i}, \mathbf{H^{SS}_{i+}}) + \left( 1-\log \mathcal{D}_{\theta 2}(\mathbf{h^{SI}_i}, \mathbf{H^{SS}_{i-}})\right)\right).
\end{split}
\end{equation}

\subsection{Disentanglement Constraint}
To preserve the unique information from the two representation spaces, a two-level disentanglement is applied. Concretely, we minimize the mutual information among the pattern vectors summarized from behavior vectors in the two domains. To achieve this, we minimize the vCLUB \citep{club} upper bound of the mutual information, which is defined as
\begin{equation}
    I_{vCLUB}(\mathbf{X; Y})=E_{p(\mathbf{X,Y})}\left[\log q_{\theta*}(\mathbf{Y|X})\right]-E_{p(\mathbf{X})}E_{p(\mathbf{Y})}\left[\log q_{\theta*}(\mathbf{Y|X})\right],
\end{equation}
where $q_{\theta*}(\mathbf{Y|X})$ is a variational distribution with parameter ${\theta*}$ to approximate $p(\mathbf{Y|X})$. %\xx{$\theta$ is used again. give an example of what the variational distribution is.}

%The unbiased estimation of vCLUB with samples $\{x_i, y_i\}$ is:
%\begin{equation}
%    \hat{I}_{vCLUB}(\mathbf{X; Y}) = \frac{1}{N^2}\sum_{i=1}^N\sum_{j=1}^N \left[\log q_\theta (y_i|x_i) - \log q_\theta (y_i|x_i) \right].
%\end{equation}

At each iteration of training, the variational approximation network trained to maximize $\log q_{\theta*} (\mathbf{Y|X})$ is first optimized, and then the main model is optimized. 

In this way, we train a vCLUB mutual information estimator for each pair of feature vectors from different domains and minimize the mutual information between each pair. The loss objective of the disentanglement module could then be formulated as 
\begin{equation}
    l_{Dis}=I_{vCLUB1}(\mathbf{H^{TT}};\mathbf{H^{SS}}) + I_{vCLUB2}(\mathbf{H^{TS}}; \mathbf{H^{ST}}).
\end{equation}

\subsection{Prediction and Training Objective}
The aggregated feature and label embeddings are then disentangled and appended to any recommendation backbone models for prediction as
\begin{equation}
    \hat{y_i}=f(\mathbf{X_i}, [\langle \mathbf{X^I_{i_k}}, y_{i_k}\rangle]_{k=1}^K, \mathbf{P^{TT}_i}, \mathbf{P^{SS}_i}, \mathbf{P^{TS}_i}, \mathbf{P^{ST}_i}), \label{eq:pred}
\end{equation}
where $f(\cdot)$ denotes an arbitrary recommendation model.

The training objective consists of the prediction loss, the sufficiency loss, and the disentanglement loss, which can be formulated as:
\begin{align}
\label{eq:loss}
l_{pred} &= \textnormal{-}\sum_{i} (y_i \log \hat{y_i} + (1- y_i) \log(1-\hat{y_i}))~ \\
\mathcal{L} &= l_{pred} + \alpha \cdot l_{Suf} + \beta \cdot l_{Dis},
\end{align}
where $\alpha$ and $\beta$ are hyperparameters to scale the loss components.

\section{Experiments}
In this section, we empirically evaluate the proposed model on three datasets. Five research questions lead the experiment part.
\begin{itemize}
    \item[\textbf{RQ1}] How does \modelname perform against the baselines?
    \item[\textbf{RQ2}] Is \modelname compatible with different backbones?
    \item[\textbf{RQ3}] Does each model component contribute to the performance? 
    %\item[\textbf{RQ4}] How does \modelname perform under the long-tail scenario?
    \item[\textbf{RQ4}] How is the efficiency of \modelname?
\end{itemize}

\subsection{Setup}
\subsubsection{Datasets}
We use three public datasets to evaluate \modelname. The statistics of the datasets are summarized in Table~\ref{tab:data}. 
\begin{table}[!h]
%\vspace{-15pt}
\caption{Dataset statistics.}
%\vspace{-5pt}
\label{tab:data}
\begin{tabular}{ccccc}
\hline
Dataset       & \# Users & \# Items & \# Samples & \# Fields \\ \hline%\# User fields & \# Item fields \\ \hline
ML-1M & $6,040$     &  $3,706$    & $970,009$     & 8 \\%5              & 3              \\
AZ-Toys       &$208,175$    & $77,687$    & $716,197$   &5 \\ %& 1              & 4              \\
ML-25M     & $162,541$    & $59,047$  & $24,187,390$   &4\\\hline%& 1              & 14  \\           \hline
\end{tabular}
%\vspace{-15pt}
\end{table}

\begin{itemize}[leftmargin=10pt]
\item\textbf{ML-1M}\footnote{\url{https://grouplens.org/datasets/movielens/1m/}} is a collection of movie ratings provided by users of the MovieLens website.
%, which contains $6,040$ users and $3,706$ movies after preprocessing and filtering. 
\item \textbf{AZ-Toys}\footnote{https://cseweb.ucsd.edu/~jmcauley/datasets.html} gathers product reviews and metadata related to toys and games available in Amazon e-commerce.
%, which contains $208,175$ users and $77,687$ items after preprocessing and filtering. 
\item \textbf{ML-25M}\footnote{https://files.grouplens.org/datasets/movielens/ml-25m.zip} is a popular movie recommendation dataset widely used in machine learning and recommender systems. %which is one of the largest publicly available datasets for movie recommendations. 
%,which has $162,541$ users and $59,047$ items after preprocessing and filtering. 
\end{itemize}

Samples with ratings greater than 3 are treated as positive samples, with the others being negative samples. The window size of the historical behaviors is 30. Data is split according to the global timestamps. Specifically, the training data lies between $[0,T_0)$, the validation data lies between $[T_0,T_1)$, and the test data lies between $[T_1,+\infty)$. The ratio of data amount for train/valid/test is $8:1:1$. %Furthermore, for all datasets, we truncated user's historical behavior sequences to be between 5 and 30 in length.  

\subsubsection{Evaluation Metrics} Two widely used metrics including AUC (Area under the ROC curve) and Log Loss (binary cross-entropy loss) are applied to evaluate the performance. 
%The area under the ROC curve (AUC) measures the overall quality of a binary classifier by calculating the area under the receiver operating characteristic (ROC) curve. Logloss measures the accuracy of a classifier's predicted probabilities compared to the true class labels, and a smaller value indicates better performance on the true data distribution. 

\subsubsection{Competing Models} We compare the proposed \modelname with the following methods: 1) conventional tabular methods including DeepFM \citep{guo2017deepfm}, DCN \citep{dcn}, PNN \citep{18pnn}, xDeepFM \citep{lian2018xdeepfm}, AutoInt \citep{autoint}, DIN \citep{18din}, DIEN \citep{19dien}, and 2) semantic-enhanced methods P5 \citep{geng2022recommendation}, UnisRec \citep{hou2022universal}, CTRL \citep{li2023ctrl}, and VQ-Rec \citep{hou2023learning}.
%, and RecFormer \citep{li2023text}.

%\noindent 1) \textbf{Conventional feature interaction mining methods:}DeepFM imposes an FM layer to replace the "wide" module in Wide\&Deep \citep{cheng2016wide} so as to model the pairwise feature interactions. DCN \citep{dcn} applied a cross network to learn both low-dimensional feature crossing and high-dimensional nonlinear feature efficiently. PNN \citep{18pnn} captures high-order feature interactions in recommendation systems by utilizing inner product operations. xDeepFM \citep{lian2018xdeepfm} combines the power of factorization machines and deep neural networks to effectively capture both linear and high-order feature interactions by introducing the cross layer. 

%\noindent 2) \textbf{User behavior modeling methods:} \textbf{DIN} dynamically models user interests and captures the sequential dependencies of user behaviors through attention mechanism. \textbf{AutoInt} enables effective modeling of high-order feature interactions by multi-head attention to select feature.

%\noindent 3) \textbf{Semantic method:} We use \textbf{CTRL} [] to compare with our network. This approach successfully integrates collaborative and semantic models by considering tabular and textual data as separate modalities and utilizes contrastive learning to align and integrate detail knowledge. Experiments shows that it outperforms previous semantic models such as \textbf{P5}[], \textbf{CTR-BERT}[] and \textbf{P-Tab}[].

\subsubsection{Implementation Details}
 We utilize Vicuna-13b \citep{vicuna2023} released by FastChat\footnote{https://github.com/lm-s} for text encoding. For a fair comparison, we fix the embedding size and the hidden layer size to be the same for all backbone models. The embedding size for the tabular domain representation is 32. The hidden layer size used for MLP is $[128,64]$. We use the bilinear networks to serve as the discriminator network in the sufficiency constraint and vCLUB mutual information estimator in the disentanglement constraint. The coefficients for the sufficiency constraint loss and disentanglement constraint loss are $0.02$ and $0.01$, respectively. For each model, the learning rate is searched in the range of $\{1e-4, 3e-4, 5e-4, 1e-3\}$, and the weight decay is searched in the range of $\{1e-5, 3e-5, 5e-5, 1e-4, 3e-4\}$. We use the Adam \citep{kingma2017adam} optimizer during training. The patience of early stop is 10. The code is available \footnote{https://github.com/KounianhuaDu/DisCo} \footnote{https://github.com/mindspore-lab/models/tree/master/research/huawei-noah/DisCo}. %\xx{how we set discriminator network in infoMAX and vCLUB mutual information estimator is also important.}
 
 %When tuning the model, the Adam[] optimizer is used with a fixed learning rate $1\times {10}^{-4} $ and weight decay $3\times {10}^{-4}$, and we tune the learning rate and weight decay for vCLUB module instead, as we found that a larger learning rate may affect the vCLUB module and break the training process. The batch size of ML-1M and AZ-Toys is set to 256, and for GoodReads it's set to 1024. Besides, we applied identical embedding layer in each backbone with embedding size 32, and for all collaborative models, the number of hidden layers L is set to 3 and the hidden size is $[128,64]$. For fair comparison, hyperparameters such as training epochs and early stop epochs are kept the same for all models, and the same tuning strategy was applied to each model.

\subsection{Overall Performance (RQ1)}
In this- section, we compare our proposed \modelname with various baseline models. The experiment results are displayed in Table~\ref{tab:main_exp}.
\begin{table*}[!t]
\centering
%\captionsetup{width=.75\textwidth}
\caption{Major results. For all the baselines, we append the user histories and their corresponding ratings/labels for fair comparison. The best result is in bold, while the second-best
value is underlined. Rel.Impr denotes the relative AUC improvement of \modelname against each baseline model. The symbol * indicates statistically significant improvement with p-value $<0.001$.}
    \label{tab:main_exp}
\begin{tabular}{cc|ccc|ccc|ccc}
\toprule
\multicolumn{2}{c|}{\multirow{2}{*}{Models}}    & \multicolumn{3}{c|}{ML-1M} & \multicolumn{3}{c|}{AZ-Toys} & \multicolumn{3}{c}{ML-25M} \\ \cline{3-11} 
\multicolumn{2}{c|}{}   & AUC & Logloss & Rel. Impr. & AUC  & Logloss  & Rel. Impr. & AUC   & Logloss  & Rel.Impr.  \\ \midrule
\multirow{7}{*}{\makecell{Tabular\\ Only\\(CRS)}} &\multicolumn{1}{|c|}{DeepFM}                 &0.7947&0.5470   &1.11\%      &0.7423&0.3720    &0.74\%       &0.8133&0.4892    &1.36\%     \\
&\multicolumn{1}{|c|}{DCN}                    &0.7961&0.5417   &0.93\%      &0.7424&0.3716    &0.73\%       &0.8134&0.4875    &1.35\%     \\
&\multicolumn{1}{|c|}{PNN}                    &0.7932&0.5451   &1.30\%      &0.7418&0.3705    &0.81\%       &0.8135&0.4869    &1.33\%     \\
&\multicolumn{1}{|c|}{xDeepFM}                &0.7938&0.5464   &1.22\%      &0.7392&0.3748    &1.16\%      &0.8093&0.4922    &1.84\%     \\
&\multicolumn{1}{|c|}{AutoInt}                &0.7945&0.5435   &1.13\%      & {0.7429}& {0.3705}    &0.66\%       &0.8128&0.4883    &1.42\%     \\
&\multicolumn{1}{|c|}{DIN}                    & {0.7976}& \underline{0.5401}   &0.74\%      &0.7424&0.3707    &0.73\%       & {0.8174}& {0.4820}    &0.86\%     \\
&\multicolumn{1}{|c|}{DIEN}                   &0.7970&0.5428   &0.82\%      &0.7446&\underline{0.3704}    &0.43\%       & {0.8189}& \underline{0.4841}    &0.68\%     \\
\midrule
\multirow{4}{*}{\makecell{Semantic\\ Enhanced}}&\multicolumn{1}{|c|}{P5}                  &   0.7937  &     0.5478     &   1.23\%         &        0.7418      &     0.3736   & 0.81\%   &    0.8091 &0.4921           &      1.90\%      \\ 
&\multicolumn{1}{|c|}{UnisRec}                  &   \underline{0.7991}  &     0.5410     &   0.55\%         &        0.7452      &     0.3837   & 0.35\%   &    0.8162 &0.5223           &      1.02\%      \\ 
&\multicolumn{1}{|c|}{CTRL}                  &   0.7979  &     0.5413     &   0.70\%         &        0.7432      &     0.3723   & 0.62\%   &    \underline{0.8189} &0.4922           &      0.68\%      \\ 
&\multicolumn{1}{|c|}{VQ-Rec}                  &   0.7972  &     0.5449     &   0.79\%         &        \underline{0.7456}      &     0.3826   & 0.30\%   &    0.8185 &0.5210           &      0.73\%      \\ 
\midrule
\multicolumn{2}{c|}{Best CRS + \modelname} &\textbf{0.8035*}&\textbf{0.5343*}   &   -         &\textbf{0.7478*}&\textbf{0.3704*}    &    -        &\textbf{0.8245*}&\textbf{0.4743*}&-         \\ 
\bottomrule 
\end{tabular}

\end{table*}

\begin{table*}[!h]
\centering
%\captionsetup{width=.75\textwidth}
\caption{Compatibility experiments. The proposed method offers additional feature fields, which can be followed by different feature interaction operations. We test its compatibility with different CTR backbones.}
    \label{tab:compatible}
\resizebox{0.99\linewidth}{!}{
\begin{tabular}{c|ccccc|ccccc|ccccc}
\toprule
\multirow{3}{*}{Backbones}  & \multicolumn{5}{c|}{ML-1M}                                                                                               & \multicolumn{5}{c|}{AZ-Toys}                                                                                                          & \multicolumn{5}{c}{ML-25M}                                                                                        \\ \cline{2-16} 
                            & \multicolumn{2}{c|}{Original}      & \multicolumn{2}{c|}{+\modelname}        & \multirow{2}{*}{\makecell{Rel.\\ Impr.}} & \multicolumn{2}{c|}{Original}                          & \multicolumn{2}{c|}{+\modelname} & \multirow{2}{*}{\makecell{Rel.\\ Impr.}} & \multicolumn{2}{c|}{Original}      & \multicolumn{2}{c|}{+\modelname} & \multirow{2}{*}{\makecell{Rel.\\ Impr.}} \\ \cline{2-5} \cline{7-10} \cline{12-15}
                            & AUC & \multicolumn{1}{c|}{LL} & \multicolumn{1}{c}{AUC} & \multicolumn{1}{c|}{LL} &                            & \multicolumn{1}{c}{AUC} & \multicolumn{1}{c|}{LL} & AUC        & \multicolumn{1}{c|}{LL}       &                            & AUC & \multicolumn{1}{c|}{LL} & AUC        & \multicolumn{1}{c|}{LL}       &                            \\ \midrule
DeepFM & 0.7947 & \multicolumn{1}{c|}{0.5470} & 0.8029 & \multicolumn{1}{c|}{0.5340} & 1.03\% & 0.7423 & \multicolumn{1}{c|}{0.3720} & 0.7462 & \multicolumn{1}{c|}{0.3708} &  0.53\% & 0.8133 & \multicolumn{1}{c|}{0.4892} & 0.8217 & \multicolumn{1}{c|}{0.4771} & 1.03\%\\
DCN & 0.7961 & \multicolumn{1}{c|}{0.5417} & 0.8035 & \multicolumn{1}{c|}{0.5343} & 0.93\% & 0.7424 & \multicolumn{1}{c|}{0.3716} & 0.7470 & \multicolumn{1}{c|}{0.3693} & 0.62\% & 0.8134 & \multicolumn{1}{c|}{0.4875} & 0.8231 & \multicolumn{1}{c|}{0.4875} & 1.19\% \\
PNN & 0.7932 & \multicolumn{1}{c|}{0.5451} & 0.8017 & \multicolumn{1}{c|}{0.5401} & 1.07\% & 0.7418 & \multicolumn{1}{c|}{0.3705} & 0.7466 & \multicolumn{1}{c|}{0.3681} & 0.65\% & 0.8135 & \multicolumn{1}{c|}{0.4869} & 0.8245 & \multicolumn{1}{c|}{0.4743} & 1.35\%\\
xDeepFM & 0.7938 & \multicolumn{1}{c|}{0.5464} & 0.7999 & \multicolumn{1}{c|}{0.5384} & 0.77\% & 0.7392 & \multicolumn{1}{c|}{0.3748} & 0.7434 & \multicolumn{1}{c|}{0.3718} & 0.57\% & 0.8093 & \multicolumn{1}{c|}{0.4922} & 0.8240 & \multicolumn{1}{c|}{0.4792} & 1.82\%\\
AutoInt & 0.7945 & \multicolumn{1}{c|}{0.5435} & 0.8029 & \multicolumn{1}{c|}{0.5343} & 1.06\% & 0.7429 & \multicolumn{1}{c|}{0.3705} & 0.7472 & \multicolumn{1}{c|}{0.3710} & 0.58\% & 0.8128 & \multicolumn{1}{c|}{0.4883} & 0.8196 & \multicolumn{1}{c|}{0.4863} & 0.83\%\\
DIN & 0.7976 & \multicolumn{1}{c|}{0.5401} & 0.8016 & \multicolumn{1}{c|}{0.5373} & 0.51\% & 0.7424 & \multicolumn{1}{c|}{0.3707} & 0.7478&\multicolumn{1}{c|}{0.3704}&0.73\%&0.8174&\multicolumn{1}{c|}{0.4820}&0.8212&\multicolumn{1}{c|}{0.4829}&0.46\%\\
DIEN & 0.7970 & \multicolumn{1}{c|}{0.5428} & 0.8025 & \multicolumn{1}{c|}{0.5353} & 0.69\% & 0.7446 & \multicolumn{1}{c|}{0.3704} & 0.7468&\multicolumn{1}{c|}{0.3693}&0.30\%&0.8189&\multicolumn{1}{c|}{0.4841}&0.8219&\multicolumn{1}{c|}{0.4792}&0.37\%\\
\bottomrule
\end{tabular}
} 

\end{table*}

From the results, one can draw the following conclusions. 
1) Our proposed \modelname can consistently outperform all the baseline models including tabular-only methods and the semantic-enhanced methods. The improvements are statistically significant under p-value $<0.001$. This shows the effectiveness of the proposed paradigm that disentangles and collaborates tabular and semantic domain knowledge for enhanced recommendation.
2) The methods that involve in semantic knowledge can surpass the conventional tabular-only methods, which shows the effectiveness of introducing external semantic knowledge into recommendation.
%that relying on semantic signals only and omitting the collaborative signals is not enough for recommendation. 
3) \modelname outperforms methods that focus on aligning the two representation spaces. For example, CTRL utilizes the contrastive learning methodology to align the two representation spaces. These methodologies tend to make representations in different representation spaces closer, where unique information is discarded during training. This validates the effectiveness of \modelname that preserves the unique information of the two different representation spaces.

\subsection{Compatibility Study (RQ2)}
Since the extracted patterns obtained from the dual-side attentive network can serve as extra features, they  could be appended to arbitrary conventional recommendation models.
In this section, we evaluate the compatibility of the proposed framework on different conventional backbones.

 The feature interaction methods of the backbones include product-based, MLP-based, and attention-based operators. We test \modelname on these different operators and justify the effectiveness of the resulting feature fields.  The results are displayed in Table~\ref{tab:compatible}. 
From the results, we can see that the proposed method could offer performance gains for various backbone models and operations. The improvements are statistically significant under p-value $<0.001$, which validates the superior compatibility of \modelname.

\subsection{Ablation Studies (RQ3)}
\subsubsection{Impact of the Dual-Side Attentive Network}
In this section, we validate the effectiveness of the proposed dual-side attentive network module. Concretely, we remove the inter-domain attention within which the two representation spaces attend to each other and only retrain the intra-domain attention. This results in the common two-tower structure in recommender systems, where the semantic and tabular representations are modeled separately for semantic dependencies and collaborative signals respectively.
While our dual-side attentive network module models both the intra-domain and inter-domain user behavior patterns.
The results of the two-tower attention and the dual-side attentive network module are displayed in Table~\ref{tab:agg}. 
\begin{table}[!h]
\centering
\caption{Experiment on the Aggregation Module.}
    \label{tab:agg}
\begin{tabular}{c|cc|cc}
\hline
\multirow{2}{*}{Datasets} & \multicolumn{2}{c|}{Two-Tower Attention}      & \multicolumn{2}{c}{Dual-Side Attention}       \\ \cline{2-5} 
 & AUC  & Logloss  & AUC  & Logloss  \\ \hline
ML-1M&0.8015&0.5360                &0.8035&0.5343                                                  \\
AZ-Toys&0.7464&0.3689&0.7478&0.3704  \\
ML-25M&0.8208&0.4837&0.8245&0.4743  \\
\hline
\end{tabular}
\end{table}

From the results, we could see that the proposed dual-side attentive network can outperform the two-tower aggregation,
which validates the effectiveness of the proposed module that captures both the intra-domain and the inter-domain knowledge. 

%We further study how the gap between the two aggregation structures changes with the number of historical behaviors changes on the movielens dataset. The result is visualized in Figure~\ref{fig:auc-logloss}. 

%\begin{figure}[!h]
%    \centering
%    \includegraphics[width=0.48\textwidth]{figures/result (3) 2.pdf}
%    \vspace{-20pt}
%    \caption{The performance of two-tower aggregation and dual-side attentive network with different behavior lengths on movielens.}
%    \label{fig:auc-logloss}
%\end{figure}

%From the result, we could see that gap between the two structure increases as the number of behaviors increases, which further validates the effectiveness of the proposed method.

\subsubsection{Impact of the Constraints}
In this section, we study the impact of the proposed constraints.
\begin{table}[h]
\centering
\caption{Impacts of the two constraints.}
    \label{tab:vclub}
    \resizebox{0.99\linewidth}{!}{
\begin{tabular}{c|cc|cc|cc}
\hline
\multirow{2}{*}{Datasets} & \multicolumn{2}{c|}{ML-1M}      & \multicolumn{2}{c}{AZ-Toys}    & \multicolumn{2}{c}{ML-25M}    \\ \cline{2-7} 
 & AUC  & Logloss  & AUC  & Logloss  & AUC  & Logloss\\ \hline
w/ Both (DisCo)&\textbf{0.8035}&\textbf{0.5343}&\textbf{0.7478}&\textbf{0.3704}&\textbf{0.8245}&\textbf{0.4743} \\
w/o Sufficiency&0.8019&0.5354&0.7474&0.3686&0.8220&0.4780 \\
w/o Disentanglent&0.8011&0.5377&0.7476&0.3687&0.8212&0.4833 \\
w/o Both&0.7988&0.5391&0.7460&0.3701&0.8163&0.4928 \\
\hline
\end{tabular}
}
\end{table}

Firstly, we conduct experiments with and without the constraints, the results of which are displayed in Table~\ref{tab:vclub}. From the results, we can see the following conclusions. 1) Both the sufficiency and the disentanglement constraints could offer performance gains to the model, with sufficiency helps to preserve task-relevant information from each space and disentanglement helps to enforce unique information from each space. 2) The two constraints collaborates and boosts performance with each other, which helps to capture both the consistency and specificity information of the two spaces.

\begin{figure}[t]
    \centering
    \vspace{-5pt}
    \includegraphics[width=0.48\textwidth,trim=60 50 55 50,clip]{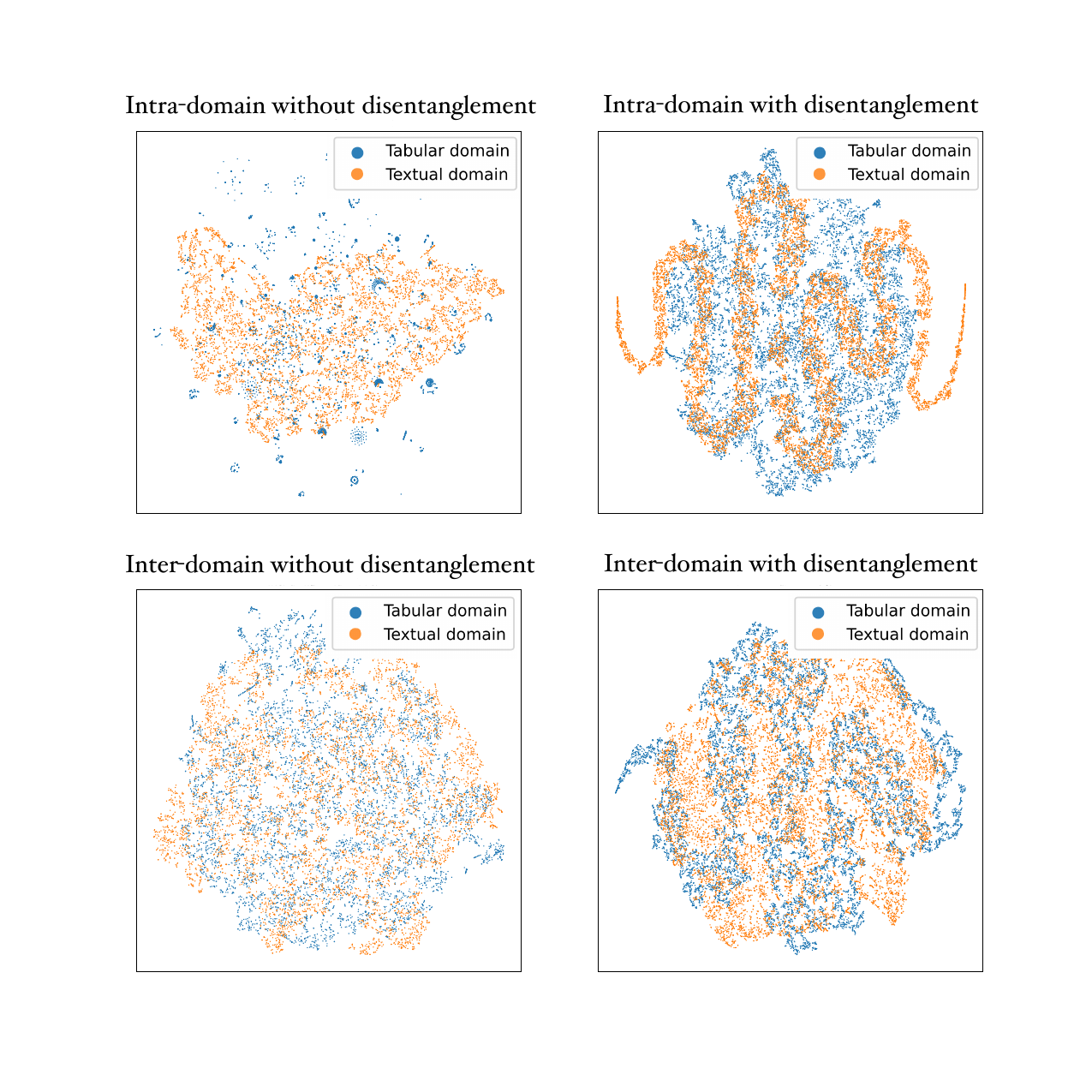}
    \vspace{-20pt}
    \caption{T-SNE visualization of the representations for the dual-side attentive network output (ML-1M).}
    \label{fig:tsne-ml}
\end{figure}

%\newpage
%\begin{figure}[t]
%    \centering
%    \vspace{-10pt}
%    \includegraphics[width=0.48\textwidth,trim=60 50 55 50,clip]{figures/visual-AZ_Toys.pdf}
%    \vspace{-20pt}
%    \caption{T-SNE visualization of the representations for the dual-side attentive network output (AZ-Toys).}
%    \label{fig:tsne-az}
%\end{figure}

In addition, we further visualize the representations of the dual-side attentive module output with and without the disentanglement constraint to dig into how the disentanglement impacts the distribution of representations. The visualization of ML-1M is displayed in Figure~\ref{fig:tsne-ml}. Visualizations of more datasets could be found in the Appendix.
%and Figure~\ref{fig:tsne-az}.

Concretely, we visualize the representations of the intra-domain pattern vectors $\mathbf{H^{TT},H^{SS}}$ and inter-domain pattern vectors $\mathbf{H^{TS},H^{ST}}$ with t-sne \citep{van2008visualizing}. The left column of Figure~\ref{fig:tsne-ml} displays the representations without the disentanglement constraint. While the right column of the figure displays those with the disentanglement constraint. From the figure, we can see that under the regularization of the disentanglement constraint, the distributions of representations from different domains are separated better and there are apparently better manifolds existing in the representation spaces with the disentanglement constraint. This illustrates that the disentanglement constraint could well separate and extract different information from the two spaces, which validates the effectiveness of our design.

%\subsubsection{Impact of Different Samples (RQ5)}
%We further study the performance of the proposed method when adopting different behavioral collection schemes. Concretely, we test two dimensions to obtain the user behaviors: temporally recent behaviors and semantic-relevant behaviors. The former truncates the most recent 30 historical behaviors for the target prediction. The latter one selects the top 30 similar historical behaviors measured by the cosine distance between the candidate item semantic embedding and the historical items embeddings. The experiment results of the study on different behavior samples are displayed in Table~\ref{tab:samples}.
%\begin{table}[!h]
%\centering
%\caption{Experiment on Different Samples.}
%    \label{tab:samples}
%\begin{tabular}{c|cc|cc}
%\hline
%\multirow{2}{*}{Datasets} & \multicolumn{2}{c|}{Temporally-Recent}      & \multicolumn{2}{c}{Semantic Neighbors}       \\ \cline{2-5} 
% & AUC  & Logloss  & AUC  & Logloss  \\ \hline
%ML-1M&0.8019&0.5354                     &0.8063&0.5338                                                   \\
%AZ-Toys&0.7478&0.3684                   &0.7482&0.3682                                               \\
%\hline
%\end{tabular}
%\end{table}

%From the results, we could see that the proposed method performs well with both the temporally-recent behaviors and semantic-relevant behaviors. In addition, obtaining behaviors with a semantic dimension retrieval could achieve a performance gain.

\subsection{Efficiency Analysis (RQ4)}
In this section, we discuss the efficiency of the proposed model. Firstly, the semantic embeddings for items could be pre-computed and stored in the indexed knowledge base, the construction of which can be done offline and only once. In addition, after the MLP used to reduce the dimension of the semantic embedding is trained, we could use the trained MLP to reduce the dimension and further keep a reduced-dimension version of the indexed knowledge base. Therefore, in the inference stage we do not need to deal with the high-dimensional semantic vectors.

\begin{table}[!h]
\centering
\caption{Training and inference time per sample (s).}
    \label{tab:time}
\begin{tabular}{c|cc|cc}
\hline
\multirow{2}{*}{Dataset} & \multicolumn{2}{c|}{ML-1M} & \multicolumn{2}{c}{AZ-Toys} \\ \cline{2-5} 
                         & Training    & Inference    & Training     & Inference     \\ \hline
DCN                      &$2.18 \times 10^{-3}$&$2.86 \times 10^{-5}$&$2.06 \times 10^{-3}$&$2.68 \times 10^{-5}$\\
DIN                      &$1.99 \times 10^{-3}$&$1.34 \times 10^{-5}$&$2.05 \times 10^{-3}$&$2.15 \times 10^{-5}$\\
%P5                       &             &              &              &               \\
%PTab                     &             &              &              &               \\
%CTR-BERT                 &             &              &              &               \\
\modelname                    &$5.04 \times 10^{-3}$&$7.42 \times 10^{-5}$&$4.29 \times 10^{-3}$&$6.89 \times 10^{-5}$\\ \hline
\end{tabular}
\end{table}

The training and inference time analysis is displayed in Table~\ref{tab:time}.  All the experiments are done on a single V100 GPU with Intel Xeon Gold 6278C 2.60GHz CPU and run three times to get the average time. %\xx{what about batch size and how we compute the inference time per sample? (I recommend the inference time per batch)}
 From the results, we can see that the proposed method does not cause a heavy overhead. 

\section{Conclusion}
Recommender systems play a vital role in our daily life. Conventional recommendation methods focus on modeling feature interactions and user behaviors within the ID-based tabular representation space and fail to capture semantic dependencies among user behaviors. Existing semantic-enhanced methods focus on aligning the tabular and semantic space, while the unique and disentangled parts of the two representation spaces are not well explored. In this paper, we propose \modelname to disentangle and collaborate the tabular and semantic representation spaces to capture both the consistent and the specific knowledge from the two spaces for enhanced recommendations. Concretely, we design three modules, namely dual-side attentive network, the sufficiency constraint, and the disentanglement constraint. To efficiently utilize the semantic knowledge, a textual description for each item is firstly obtained and encoded by LLMs, the embedding of which is then stored into an indexed knowledge. The dual-side attention module models intra-domain and inter-domain patterns to offer additional knowledge for arbitrary recommendation backbones, which is constrained by the designed sufficiency and disentanglement constraints. The two constraints force the model to preserve useful information and extract unique information from the two spaces. Extensive experiments and ablation studies on three datasets and various backbone models justify the effectiveness of the proposed method.  
%\ljh{if possible, give future work}

%\begin{acks}
%To Robert, for the bagels and explaining CMYK and color spaces.
%\end{acks}

\bibliographystyle{ACM-Reference-Format}
\bibliography{sample-base}

\newpage
\appendix

\section{Visualizations}
\begin{figure}[h]
    \centering
    \vspace{-10pt}
    \includegraphics[width=0.48\textwidth,trim=60 50 55 50,clip]{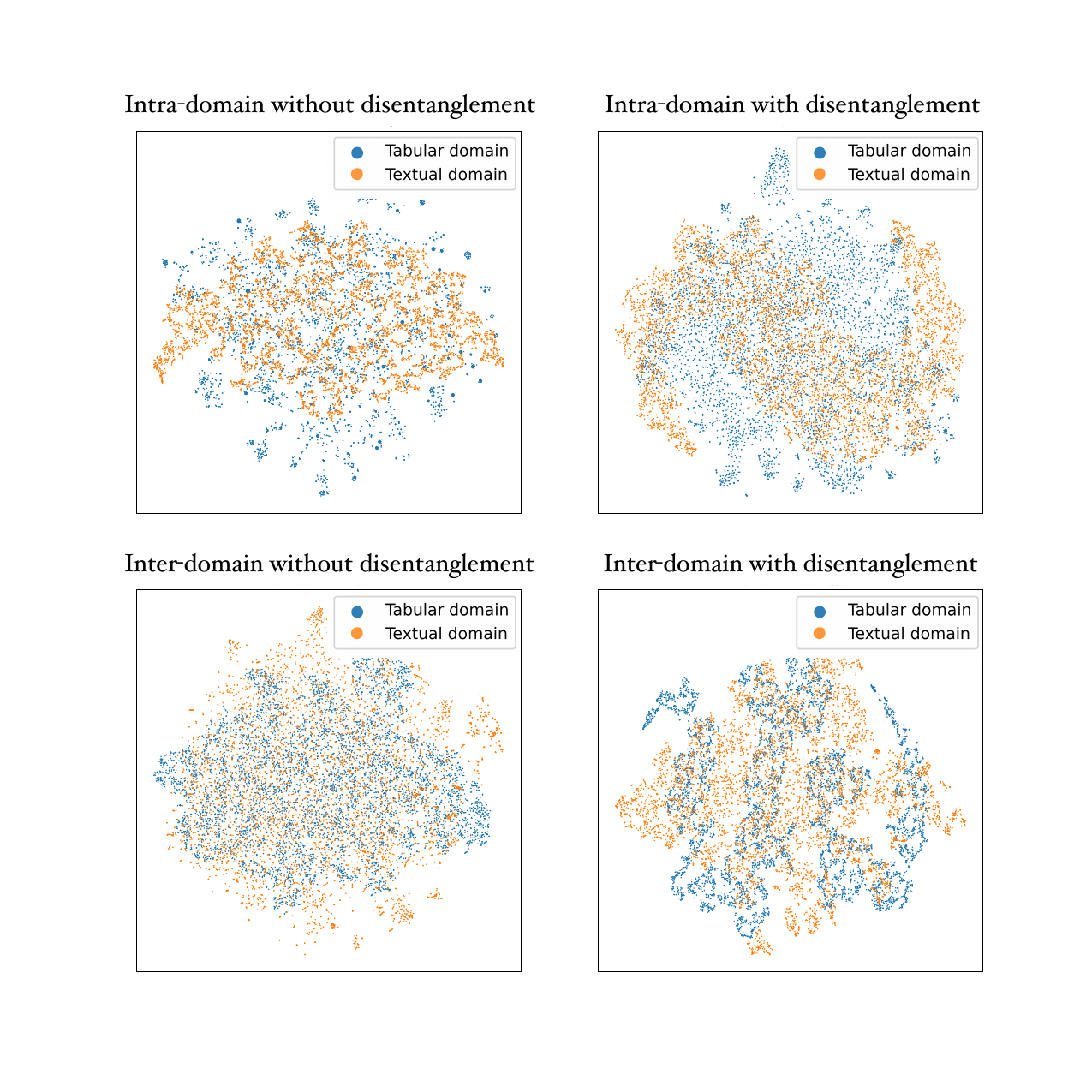}
    \vspace{-20pt}
    \caption{T-SNE visualization of the representations for the dual-side attentive network output (AZ-Toys).}
    \label{fig:tsne-az}
\end{figure}

\begin{figure}[h]
    \centering
    \vspace{-5pt}
    \includegraphics[width=0.48\textwidth,trim=60 50 55 50,clip]{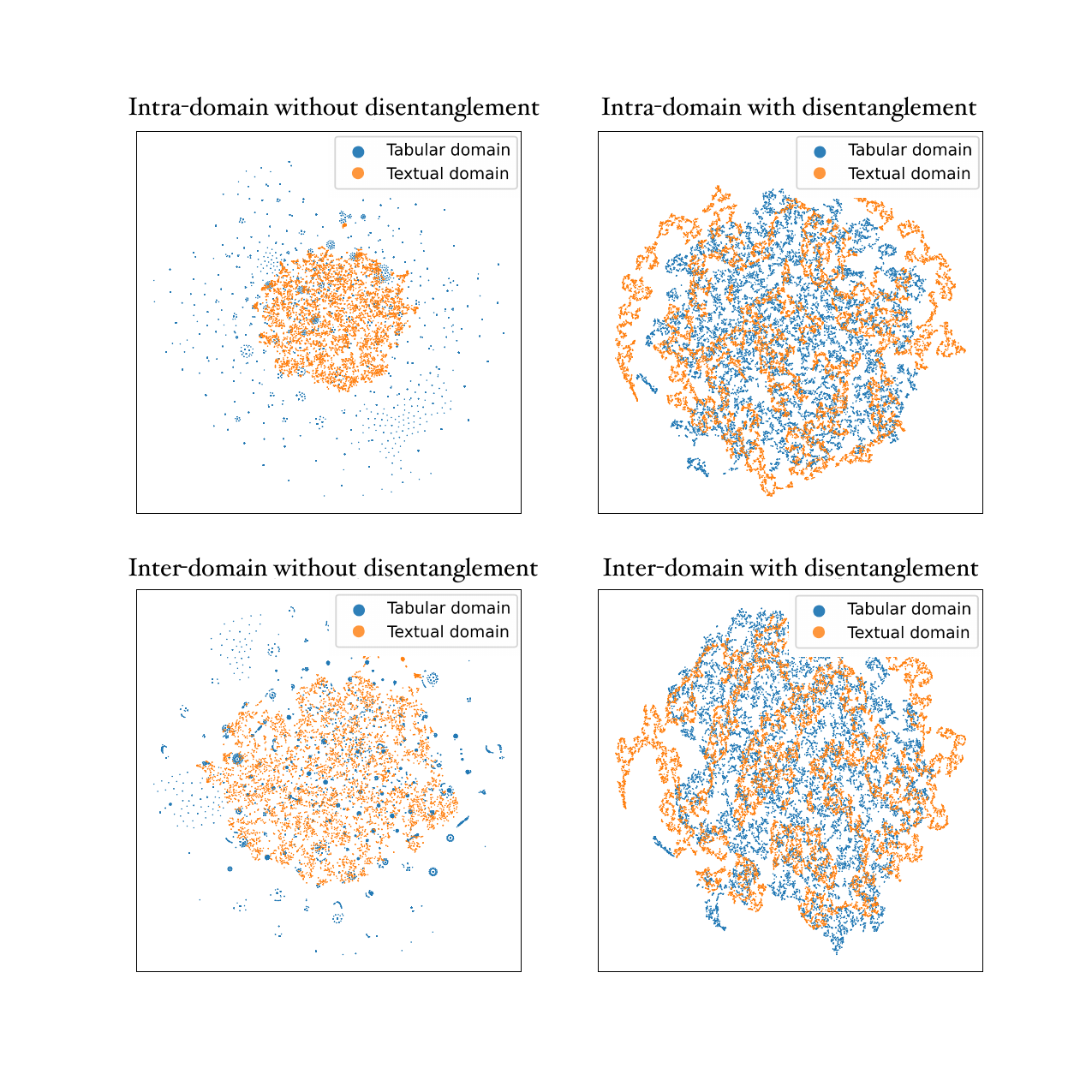}
    \vspace{-20pt}
    \caption{T-SNE visualization of the representations for the dual-side attentive network output (ML-25M).}
    \label{fig:tsne-ml25}
\end{figure}

Visualizations of the representations for the dual-side attentive network on AZ-Toys and ML-25M are displayed in Figure~\ref{fig:tsne-az} and Figure~\ref{fig:tsne-ml25}.

\begin{table*}[!h]
\centering
\caption{Items used in the case study.}
    \label{tab:case1}
\begin{tabular}{c|cc}
\hline
Item Title & Galaxy Quest & Babe: Pig in the City \\\hline
Genre provided in the movielens dataset   & Adventure  & Children's movie\\\hline
\makecell{Tags provided by Douban \\ (\textbf{Open world knowledge, unknown by the tabular model})}   & Comedy, Action, Fantasy and adventure.& Comedy,  Fantasy and Adventure.  \\
\hline
\end{tabular}
\end{table*}

\begin{table*}[!h]
\centering
\caption{Similarities of items in different spaces.}
    \label{tab:case2}
\begin{tabular}{c|ccc}
\hline
Cosine Similarity & In Tabular Space & In Semantic Space (BERT)& In Semantic Space (Vicuna-13b)\\\hline
Galaxy Quest --  Babe: Pig in the City & -0.0210                     & 0.0599  & 0.9606 \\\hline
\end{tabular}
\end{table*}
\section{Experiments on Long-Tail Data}
In this section, we validate the effectiveness of \modelname on the long-tail data where features are less-hit in the tabular representation space. Concretely, we sort items based on their frequency of occurrences
in the training set. The bottom 10\% in terms of frequency are classified as long-tail items. Then we study the performance of the best-performed tabular-only model DIN and that with the proposed DisCo model. 

\begin{table}[!h]
\centering
\caption{Experiment on the tail Data.}
    \label{tab:longtail}
\begin{tabular}{c|cc|cc}
\hline
\multirow{2}{*}{Datasets} & \multicolumn{2}{c|}{DIN}      & \multicolumn{2}{c}{DIN + \modelname}       \\ \cline{2-5} 
 & AUC  & Logloss  & AUC  & Logloss  \\ \hline
ML-1M&0.6710&0.6564 &0.6934&0.6308   \\
AZ-Toys&0.6673&0.3982&0.7416&0.3778 \\
ML-25M&0.7963&0.5430&0.8032&0.5380  \\
\hline
\end{tabular}
\end{table}

The performance comparisons are displayed in Table~\ref{tab:longtail}. From the results, we can see that the proposed method gives a significant performance boost on the tail data. Since the general knowledge contained in the pretrained semantic embeddings helps to complement the less-trained features in the tabular representation space, where the semantic information is injected through the dual-side attentive network under the regularizations of the constraints.

\section{Case Study}
In this section, We would like to provide an example to support the claim that encodings from LLMs can capture open-world knowledge as below.

Given two items in the movielens dataset: "Galaxy Quest" and "Babe: Pig in the City". Their genres provided in the dataset and the tags provided by open world are listed in Table~\ref{tab:case1}. 

For the two items:
\begin{itemize}
    \item The two items do not share any common tokens.
    \item The two items do not share any common features given in the dataset.
    \item But they are actually close as they share a lot common features in the open world (e.g., the tags given by Douban).
\end{itemize}

We then compute the cosine similarities between the above two items in the tabular space and that in the semantic space as in Table~\ref{tab:case2}. (Note that no generation is involved. We use the same genres provided by the movielens dataset to obtain the item encodings for both the tabular encodings and the semantic encodings.)

From the results, we can see that 
\begin{itemize}
    \item Tabular representations cannot find the relevance between the two items, since no common features exist.
    \item Semantic representations by small language models cannot well find the relevance between the two items, since no common tokens exist.
    \item Semantic representations by LLMs can well find the relevance between the two items, even there are nearly no common tokens shared between the two items in the dataset. Since there are many anchors about the two items existing in the large open world training corpus (e.g., the common tags given by Douban, the similar descriptions of movies, etc.), as they are trained together, the representations of the two items tend to get close.
\end{itemize}

The above case study can justify that the encodings from LLMs can help to capture open-world knowledge.
%\xx{Too short. should give an example and explain what makes DisCo perform well on long-tail data.}


%Lorem ipsum dolor sit amet, consectetur adipiscing elit. Morbi
%malesuada, quam in pulvinar varius, metus nunc fermentum urna, id
%sollicitudin purus odio sit amet enim. Aliquam ullamcorper eu ipsum
%vel mollis. Curabitur quis dictum nisl. Phasellus vel semper risus, et
%lacinia dolor. Integer ultricies commodo sem nec semper.


%Etiam commodo feugiat nisl pulvinar pellentesque. Etiam auctor sodales
%ligula, non varius nibh pulvinar semper. Suspendisse nec lectus non
%ipsum convallis congue hendrerit vitae sapien. Donec at laoreet
%eros. Vivamus non purus placerat, scelerisque diam eu, cursus
%ante. Etiam aliquam tortor auctor efficitur mattis.


%Nam id fermentum dui. Suspendisse sagittis tortor a nulla mollis, in
%pulvinar ex pretium. Sed interdum orci quis metus euismod, et sagittis
%enim maximus. Vestibulum gravida massa ut felis suscipit
%congue. Quisque mattis elit a risus ultrices commodo venenatis eget
%dui. Etiam sagittis eleifend elementum.

%Nam interdum magna at lectus dignissim, ac dignissim lorem
%rhoncus. Maecenas eu arcu ac neque placerat aliquam. Nunc pulvinar
%massa et mattis lacinia.

\end{document}